\documentclass[floatfix,
 reprint,
superscriptaddress,
nobibnotes,
 amsmath,amssymb,
 aps,
]{revtex4-2}

\usepackage{graphicx}% Include figure files
\usepackage{dcolumn}% Align table columns on decimal point
\usepackage{bm}% bold math
\usepackage{natbib}
\begin{document}

\preprint{APS/123-QED}

\title{Breaking electron pairs in the pseudogap regime of \\ SmTiO$_3$/SrTiO$_3$/SmTiO$_3$ quantum wells}% Force line breaks with \\

\author{Xinyi Wu}
\author{Lu Chen}
\author{Arthur Li}
\author{Megan Briggeman}
\affiliation{ 
Department of Physics and Astronomy, University of Pittsburgh, Pittsburgh, Pennsylvania 15260, United States}
\affiliation{ 
Pittsburgh Quantum Institute, Pittsburgh, Pennsylvania 15260, United States.}

\author{Patrick B. Marshall}
\author{Susanne Stemmer}
\affiliation{%
Materials Department, University of California, Santa Barbara, California 93106-5050, United States}

\author{Patrick Irvin}
\author{Jeremy Levy}
\email{jlevy@pitt.edu.}
\affiliation{ 
Department of Physics and Astronomy, University of Pittsburgh, Pittsburgh, Pennsylvania 15260, United States}
\affiliation{ 
Pittsburgh Quantum Institute, Pittsburgh, Pennsylvania 15260, United States.}

\date{\today}% It is always \today, today,

\begin{abstract}
The strongly correlated two-dimensional electron liquid within SmTiO$_3$/SrTiO$_3$/SmTiO$_3$ quantum well structures exhibits a  pseudogap phase when the quantum well width is sufficiently narrow. Using low-temperature transport and optical experiments that drive the quantum-well system out of equilibrium, we find evidence of mobile, long-lived, negatively-charged quasiparticles, consistent with the idea that the pseudogap phase arises from strongly bound electron pairs. 
\begin{description}
\item[Keywords]
Correlated electronics; complex-oxide heterostructure; quantum wells; pseudogap phase
\end{description}
\end{abstract}

\maketitle

%\tableofcontents

%\section{Introduction}

\textit{Introduction.---}The pseudogap phase is an unusual state of electronic matter which is primarily associated with unconventional superconductivity \cite{Ding1996-vm,Loeser1996-jo,Timusk1999-pr,Lee2006-os,Shimojima2014-qt}.  It is characterized by a reduction in spectral weight near the Fermi energy,
% associated with an excitation gap for fermionic excitations, 
and generally appears above the superconducting transition temperature $T_c$ in the low-doped regime. Understanding the pseudogap state has been the subject of intense focus for cuprate superconductors as it provides unique insights into correlated electron physics. Apart from the cuprates, pseudogap behavior has also been found in iron-based superconductors \cite{Tohyama2011-bf}, heavy-fermion compounds \cite{Sidorov2002-pi}, charge-density-wave systems \cite{Rossnagel2011-ad}, rare-earth nickelates \cite{Allen2015-zr},  ultracold atomic gas systems \cite{Stewart2008-zj,Jochim2003-bz}, and SrTiO$_3$-based heterostructures \cite{Richter2013-lr} and quantum wells \cite{Marshall2015-cv}. 
 
A long-standing question for pseudogap phenomena in superconductors concerns its physical origin, specifically, whether it arises to pre-formed Cooper pairs or some other mechanism \cite{Bozovic2020-tt}.  Many theoretical descriptions of the pseudogap phase take place within the the BCS-BEC crossover picture \cite{Randeria2014-fz}.  Alternative explanations involve proximity to competing phases, which are prevalent in cuprate superconductors \cite{He2011-qj}.

%fluctuations can give rise to pairing and at the same time suppress superconductivity \cite{Hetel2007-qj}. Disorder can also play a role in strongly correlated systems and has been shown in theory to induce a zero-energy anomaly \cite{Altshuler1979-ah} and stabilize a pseudogap state \cite{Lee1999-aw}. 
 
SrTiO$_3$ is one of a few semiconductors known to exhibit superconducting behaviour \cite{Schooley1964-xx} with a dome-like shape of $T_c$ as a function of electron density \cite{Koonce1967-sl}. It is considered to be an unconventional superconductor with an unknown pairing mechanism. Recent advances in understanding the nature of superconductivity in SrTiO$_3$ arose from SrTiO$_3$-based heterostructures~\cite{Pai2018-br}, in particular  LaAlO$_3$/SrTiO$_3$  \cite{Reyren2007-sz,Caviglia2008-bf}, which exhibit gate-tunable superconductivity through a quantum critical point, and pseudogap behavior past the low-density boundary of the superconducting dome \cite{Richter2013-lr}.  Low-dimensional structures formed at the LaAlO$_3$/SrTiO$_3$ interface show direct evidence for pre-formed electron pairs in confined single-electron transistor geometries \cite{Cheng2015-fv,Prawiroatmodjo2017-ke} and extended electron waveguide geometries \cite{Annadi2018-az,Briggeman2020-cl}.

\begin{figure*}
\includegraphics[width=15cm]{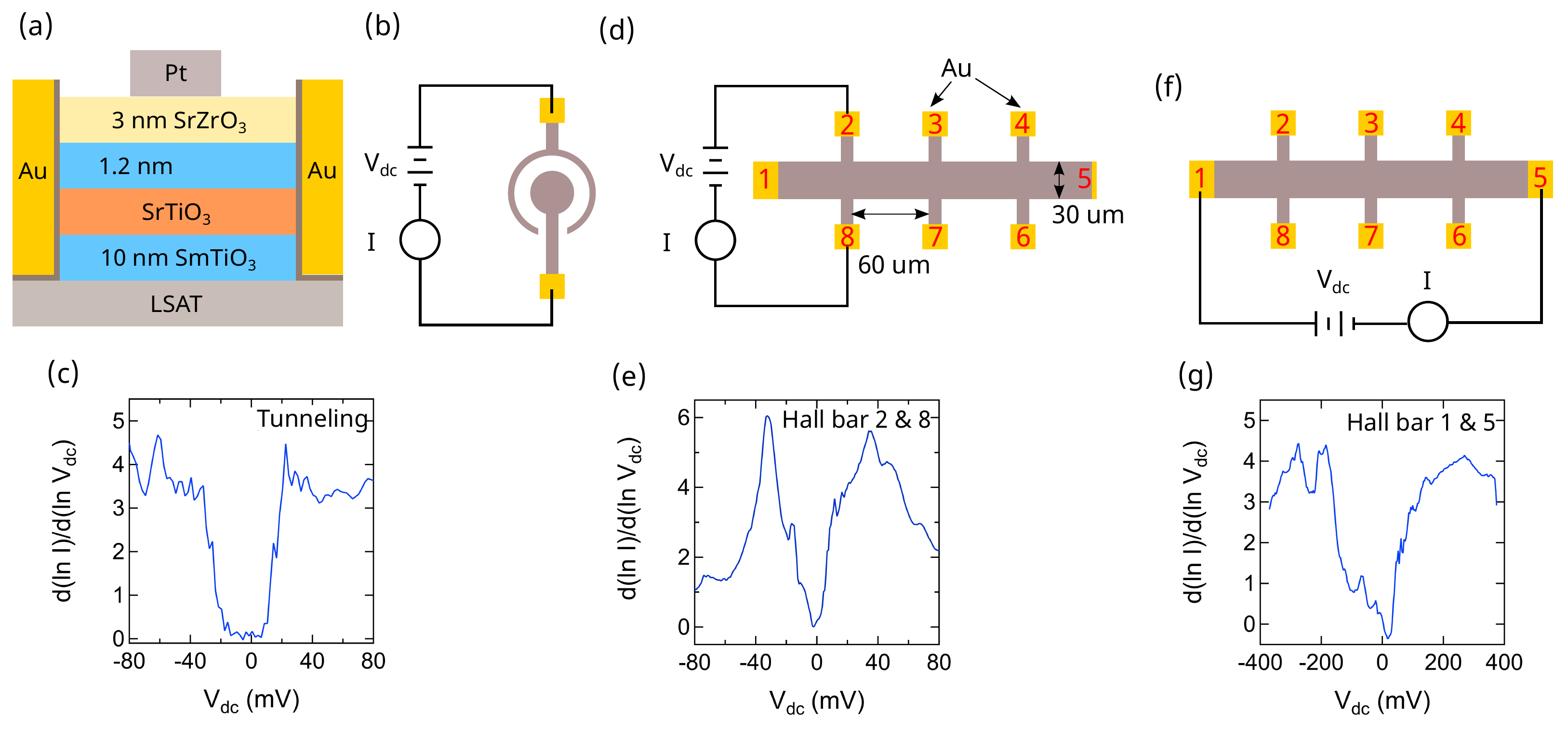}
\caption{\label{fig:diagram}
(a) Cross-sectional schematic of the SmTiO$_3$/SrTiO$_3$/SmTiO$_3$ quantum well structure. The SrTiO$_3$ quantum well contains two SrO layers. Ti (grey) and Au (yellow) contacts are made to be in direct contact with the conductive SrTiO$_3$ layer.
(b) Schematic of the tunneling device with the experimental setup for two-terminal $I$-$V$ measurements between the top and interface electrodes, separated by a radial distance of 60 $\mu$m.
(c) Normalized conductance spectra (d($\ln I$)/d($\ln V$) vs $V$)  of the tunneling device measured at $T=2$ K. A pseudogap feature with a characteristic onset voltage of $V_T=27$ mV.
(d) Schematic of SmTiO$_3$/SrTiO$_3$/SmTiO$_3$ channel device configured for two-terminal current-voltage measurements between electrodes 2 and 8, separated at a distance of 90 $\mu$m. 
(e) Normalized conductance spectra shows a reduction of conductance near zero voltage and  a characteristic onset at $V_T=40$ mV.
(f) Schematic of SmTiO$_3$/SrTiO$_3$/SmTiO$_3$ channel device and two-terminal current-voltage measurements between electrodes 1 and 5, separated at a distance of 240 $\mu$m. 
(g) Normalized conductance spectra shows a reduction of conductance near zero voltage and a characteristic onset at $V_T=120$ mV.
}
\end{figure*}

The hybrid molecular beam epitaxy (hybrid-MBE) growth technique has enabled new families of SrTiO$_3$-based quantum wells to be designed with high doping density and ultranarrow widths.  Quantum well systems containing rare-earth (Re) titanate ReTiO$_3$/SrTiO$_3$ interfaces demonstrate a variety of emerging properties that can be tuned by the quantum well width, including metal-to-insulator transition \cite{Jackson2013-bo}, magnetism \cite{Moetakef2012-sr}, quantum critical behaviour \cite{Zhang2014-qy,Mikheev2015-yw}, and a separation of transport and Hall scattering rates \cite{Mikheev2015-yw}. 
Narrow SrTiO$_3$ quantum wells surrounded by antiferromagnetic SmTiO$_3$ barriers possess quantum well electron densities as high as $n_{2D}=7\times10^{14}$ cm$^{-2}$.  
The narrowest quantum well structures show evidence of non-Fermi-liquid behavior \cite{Jackson2014-tm,Stemmer2018-hj},
%and a suppressed single-particle density of states (DOS) at the Fermi energy with sharp coherence peaks above a appearing below $T=10~$K, 
and tunneling experiments yield telltale characteristics of a pseudogap phase \cite{Marshall2016-jd}. Pseudogap behavior is observed with antiferromagnetic SmTiO$_3$ barriers and not ferrimagnetic GdTiO$_3$ barriers, which provide some evidence for a spin-density-wave phase.  A metal-insulator transition has also been observed at single SmTiO$_3$/SrTiO$_3$ heterointerfaces at a temperature $T$~110 K \cite{Ahadi2017-hq}. Recent angle-resolved photoemission experiments have correlated the pseudogap regime in ultranarrow SmTiO$_3$ quantum wells with a change in Fermi surface topology, i.e., a Lifshitz transition.  A similar connection is known to exist between the Lifshitz transition in the LaAlO$_3$/SrTiO$_3$ system \cite{Joshua2012-vy}.  Quantum confinement is exceptionally strong in the narrowest SmTiO$_3$ quantum wells, which could lead to a significant increase in pairing strength or stabilization of some competing phase.

\begin{figure}
\includegraphics[width=5.5cm]{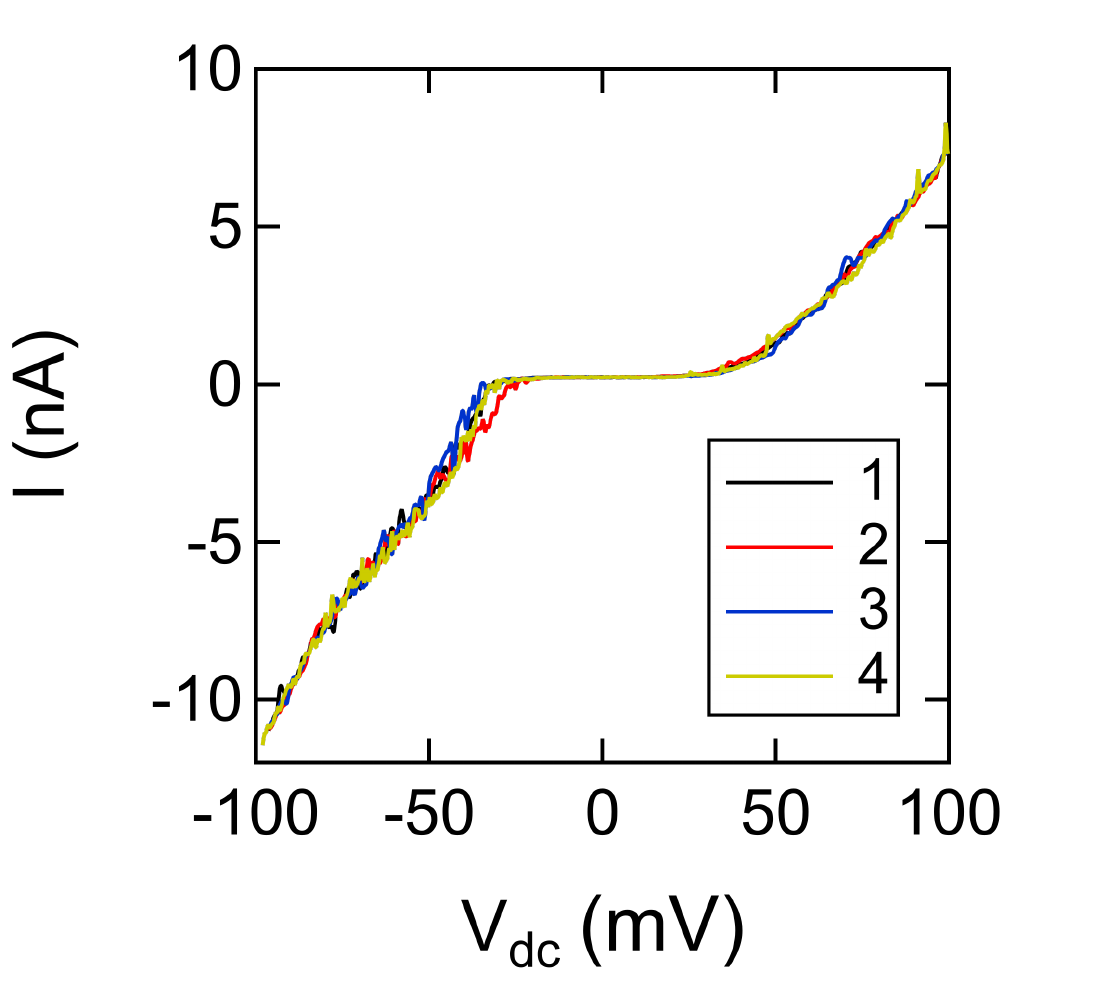}% Here is how to import EPS art
\caption{\label{fig:fluctuation} Four $I$-$V$ curves taken consecutively under the same configuration shown in Fig.~\ref{fig:diagram}(d).  Fluctuations in and around the threshold for conduction are apparent.}
\end{figure}

\textit{Experiments.---}  Our aim is to explore a possible connection between low-temperature pseudogap behavior in LaAlO$_3$/SrTiO$_3$ and higher-temperature pseudogap behavior in SmTiO$_3$/SrTiO$_3$/SmTiO$_3$ quantum wells.  The hypothesis is that the pseudogap phase in SmTiO$_3$/SrTiO$_3$/SmTiO$_3$ arises due to an electron pairing mechanism that already exists in SrTiO$_3$ and is somehow enhanced by quantum confinement.  

To test this hypothesis, we have designed low-temperature transport and optical experiments designed to break down the pseudogap phase, i.e., break electron pairs. Specifically, we investigate the transport behavior of SrTiO$_3$ quantum well when the system is driven out of equilibrium using two types of disturbances: (1) voltages exceeding the pseudogap energy and (2) optical excitation. 

Fig.~\ref{fig:diagram}(a) shows a schematic cross-section of the quantum well structure. SrZrO$_3$/SmTiO$_3$/SrTiO$_3$/SmTiO$_3$ films are grown on (La$_{0.3}$
Sr$_{0.7}$)(Al$_{0.65}$Ta$_{0.35}$)O$_3$ substrates using hybrid MBE, as detailed in \textcite{Moetakef2012-sr,Zhang2014-qy}. SrZrO$_3$ is deposited between the top SmTiO$_3$ layer and Pt top-gate electrodes as a tunneling barrier for top-gated tunneling devices. 
Electrical contact to the SrTiO$_3$ is achieved by Ar-ion etching (60$~$nm) followed by Ti/Au deposition ($4~$nm/$50~$nm).
Both SrTiO$_3$/SmTiO$_3$ interfaces  contribute $\sim$ 3.4$\times$10$^{14} \textrm{ cm}^{-2}$ electrons to the quantum well, leading to an expected total doping density of $\sim 7\times 10^{14} \textrm{ cm}^{-2}$  \cite{Moetakef2012-sr, Mikheev2015-yw}.

In order to probe  transport properties,
the quantum well heterostructure is patterned into vertical and lateral transport structures (Figs.~\ref{fig:diagram}(b,d)).  
Current-voltage (I-V) measurements are performed between several pairs of contacts in channels in a dilution refrigerator capable of measuring electrical properties at temperatures ranging from 50 mK to 300 K. Differential conductance spectra d$I$/d$V$ vs $V$ and normalized conductance $d(\ln I)/d(\ln V)=(dI/dV)(I/V)^{-1}$ \cite{Lang1986-my} are obtained via numerical differentiation.

In both vertical and horizontal geometries, the normalized conductance at $T$= 50 mK   (Fig.~\ref{fig:diagram}(c,e,g)) shows a suppression around zero bias, consistent with previous reports of pseudogap behavior~\cite{Marshall2016-jd}.  Here we use the local maxima of the coherence peaks in the vertical geometry to estimate a pseudogap 
energy $\Delta=eV_T=27$~meV, based on Fig.~\ref{fig:diagram}(c).  In both vertical and lateral geometries, there are portions of the circuit in which carriers move in the plane, which complicates the analysis.  Generally, the measured pseudogap is expected to over-estimate the intrinsic gap due to voltage drops in the plane. 
Other reported signatures, e.g., current fluctuations near the threshold for conduction (Fig.~\ref{fig:fluctuation}), are also observed.  Similar potential fluctuations were reported by \textcite{Hardy2017-od}.

\begin{figure}
\includegraphics[width=6.5cm]{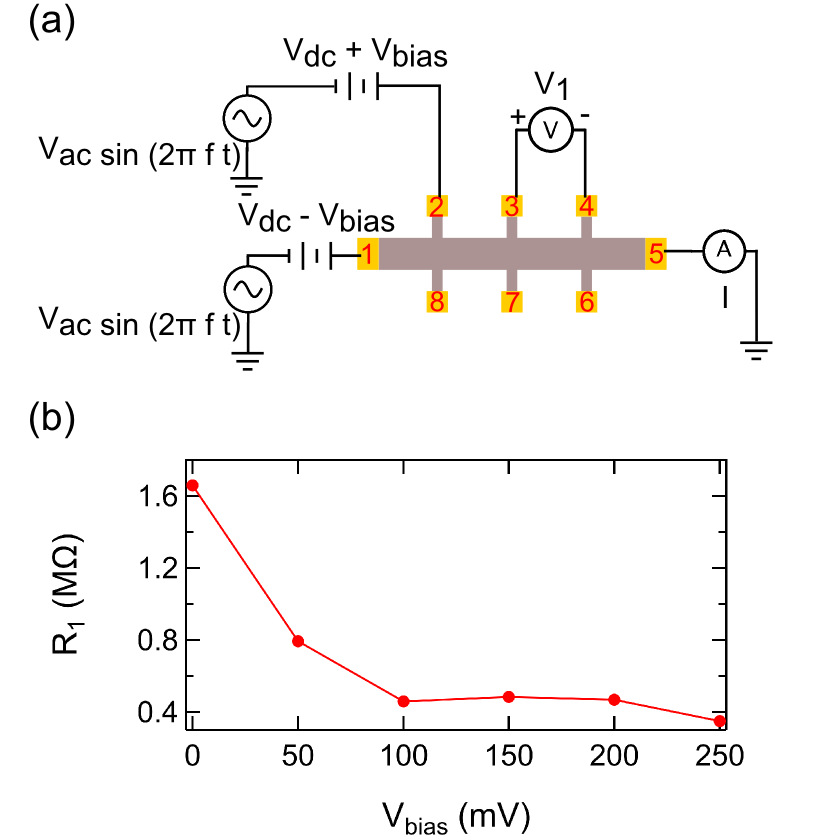}% Here is how to import EPS art
\caption{\label{fig:7}(a) Set up for the dc voltage excitation experiments. A voltage bias $V_{bias}$ is applied between two contacts on contacts labeled 1 and 2 while four-terminal lock-in measurements are performed between contacts 3 and 4. (b) Four-terminal resistance $R = V_1/I$ as a function of $V_{bias}$.}
\end{figure}

Non-local non-equilibrium four-terminal conductance measurements are conducted using the setup shown in Fig.~\ref{fig:7}(a).  At one of the channel, a combination of differentially applied dc voltage $V_{bias}$ and common-mode dc+ac voltage $V_{dc}+V_{ac}$sin(2$\pi ft)$ is applied, while the resulting ac voltage $V_1$ and current $I$.  The resistance $R_1=dV_1/dI$ (Fig.~\ref{fig:7}(b)) for $V_{ac}$ = 20 mV, frequency $f$ = 13 Hz, and $V_{dc}$ = -40 mV. $R_1$ decreases by a factor of four  as $V_{bias}$ is increased from 0 mV to 100 mV, and is relative stable for 100 mV $< V_{bias}$.  This characteristic voltage is comparable to the characteristic pseudogap voltage $V_T$ measured in Fig.~\ref{fig:diagram}(f,g).
The quantum well system is driven out of equilibrium by a differentially applied voltage $V_{bias}$, which induces a current between the two electrodes, while the resulting change in resistance along the main channel is measured ``downstream`` by driving with the combination of a sinosoidal voltage source $V_{ac}$ and a low dc bias $V_{dc}$ and monitoring the resulting current $I$ at electrode 5 and the voltage drop $V_1$ between electrodes 3 and 4. The measurements are non-local because changes in four-terminal conductance are influenced by an excitation in a region outside of where $V_1$ is measured. 

\begin{figure} [!htb]
    \centering \includegraphics[width=8.5cm]{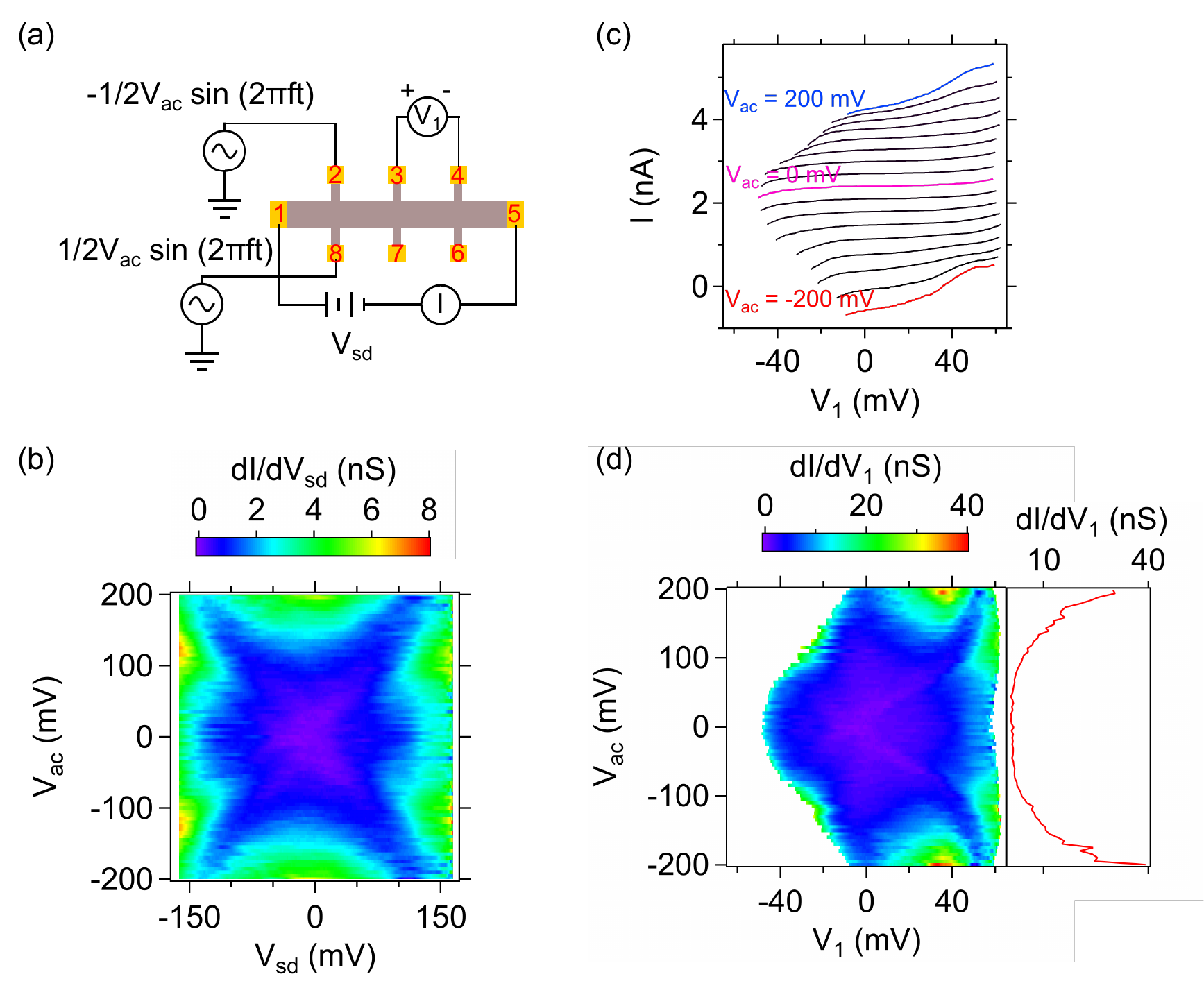}
    \caption{\label{fig:nonlocal-Hall} Transport measurements under non-equilibrium excitation. (a) Experimental setup. AC driving with magnitude $\frac{1}{2}V_{ac}$ and frequency $f$~= 1000 Hz are applied differentially to two electrodes labeled 2 and 8 on the left-hand side of the channel device. A quasi-dc voltage $V_{sd}$ is applied between electrodes 1 and 5, swept between -160 mV and 160 mV.  (b) Intensity plot of two-terminal conductance versus $V_{sd}$ and $V_{ac}$.  (c) Measured four-terminal conductance in the region between electrodes 3 and 4 for various $V_{ac}$.  Curves are offset for clarity. (d) Four-terminal conductance intensity map versus $V_1$ and $V_{ac}$. (e) Linecut of four-terminal conductance at $V_{sd}=0$~V.  
    }
\end{figure} 

We expand our transport measurements in the channel structure using an experimental setup illustrated in
Fig.~\ref{fig:nonlocal-Hall}(a).  
The left side of the channel (electrodes 2 and 8) is driven driven differentially with amplitude $\frac{1}{2}V_{ac}$ and frequency $f$~= 1000 Hz, resulting in a net voltage driving across the channel with amplitude $V_{ac}$.  At the same time, a quasi-dc voltage $V_{sd}$ is applied between electrodes 1 and 5, with current $I$ measured at electrode 5 and voltage $V_1$ measured between electrodes 3 and 4 ($V_1$). The two-terminal conductance $dI/dV_{sd}$ and four-terminal conductance $dI/dV_1$ on the right side of the device are calculated by numerical differentiation as a function of $V_{ac}$ excitation on the left side of the device. 

The four-terminal $I$-$V_1$ curves are shown in Fig.~\ref{fig:nonlocal-Hall}(b) for various values of $V_{ac}$.  Curves are offset for clarity. 
A noticeable asymmetry exists in $I(V_1)$ with respect to $V_1$, in that the magnitude of the current at negative $V_1$ exceeds the value at positive $V_1$, i.e., $|I(-V_1<0)|>|I(V_1>0)|$. This asymmetry increases as $|V_{ac}|$ is increased.   
The two-terminal and four-terminal differential conductances $dI/dV_{sd}$ (Fig.~\ref{fig:nonlocal-Hall}(c,d)) also show some notable similarities and differences.  The parameter range $V_{sd}\approx0; V_{ac}\approx0$ exhibits a low conductance, consistent with the pseudogap behavior shown in Fig.~\ref{fig:diagram}(f).  The voltage threshold for conductance $dI/dV_{sd}$ initially decreases monotonically with $V_{ac}$, and then begins to increase again.  The resulting threshold voltage contour is approximately ellipsoidal with characteristic lobes of decreased conductance when $V_{ac}$ and $V_{sd}$ coincide in magnitude.

The four-terminal conductance $dI/dV_1$ (Fig.~\ref{fig:nonlocal-Hall}(d)) is decidedly asymmetric, showing a much larger threshold for conduction when $V_1>$0 than when $V_1<$0.  The asymmetry becomes more pronounced when $V_{ac}$ increases.  For example, when $V_{ac}$=100 mV, the positive threshold for conduction is approximately $V_1$=50 mV, while the negative threshold is approximately $V_1$=-20 mV.

\begin{figure}
\includegraphics[width=5.8cm]{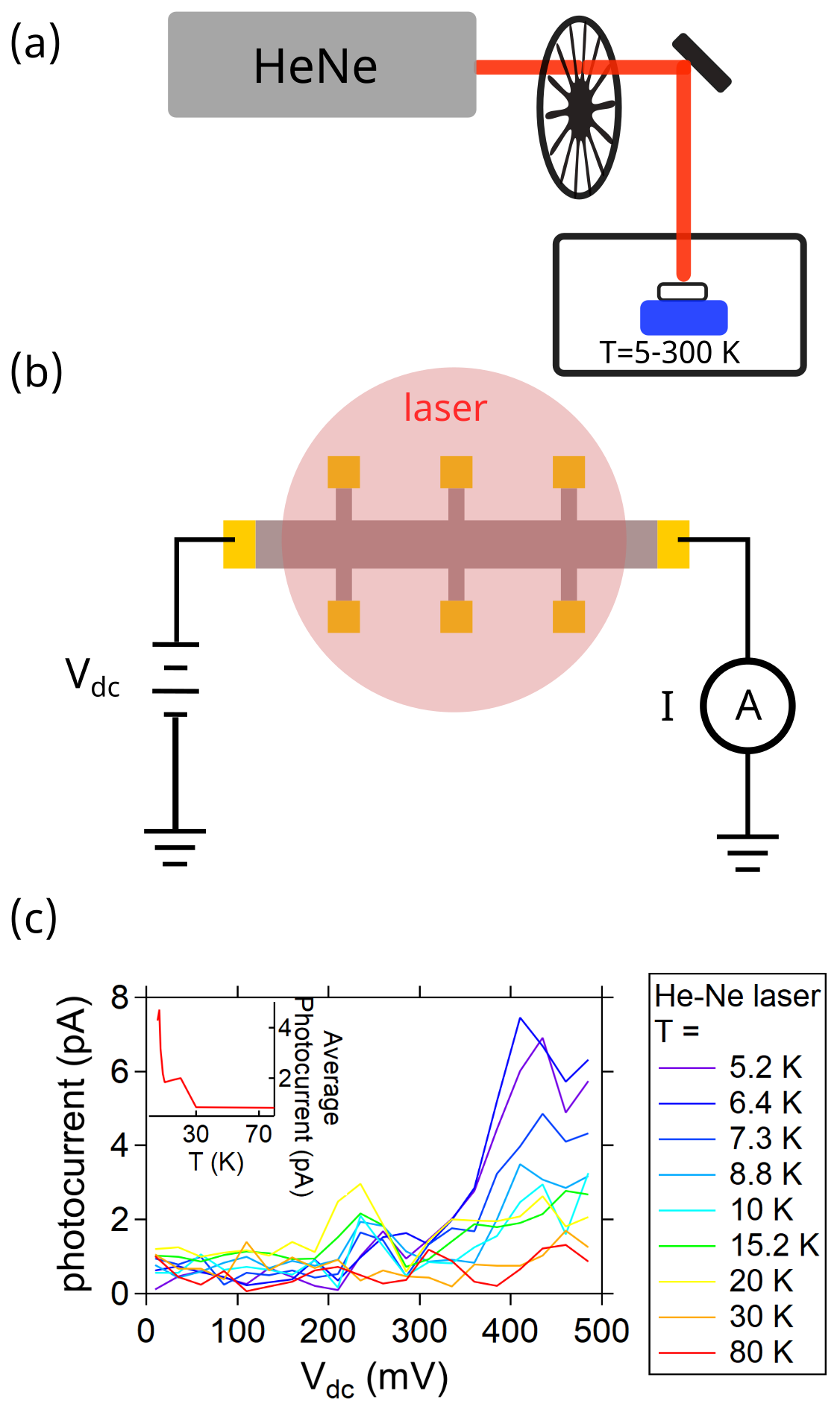}% Here is how to import EPS art
\caption{\label{fig:optical} (a) Optical setup for the two-terminal photocurrent measurements.  
(b) Photo-induced current changes $\Delta I$ are measured as a function of $V_{dc}$ and temperature.  Photocurrent is observed when $V_{dc}$ exceeds a threshold $\approx 300$~mV, and for temperatures $T<$30 K.  Inset graph shows average photocurrent over $V_{dc}$=400-500~mV.}
\end{figure}

We now examine the effect of optical illumination on transport properties of the SmTiO$_3$/SrTiO$_3$/SmTiO$_3$ channel.  Photo-induced current changes in the SmTiO$_3$/SrTiO$_3$/SmTiO$_3$ quantum wells are investigated by illuminating the channel with light from a continuous-wave HeNe laser ($\lambda=632.8$~nm, intensity $I=1\times 10^6$ W/m$^2$), modulated at frequency $f=247$~Hz using a mechanical chopper (Fig.~\ref{fig:optical}(a)).  Photo-induced current changes $\Delta I$ are measured versus the dc bias across the channel $V_{dc}$ (Fig.~\ref{fig:optical}(b)). The sample temperature is controlled by an optical cryostat which enables measurements to be conducted between $T=5.2$~K and $T=80$~K.  Non-negligible photocurrent is observed above a voltage threshold of approximately 300 mV, and for temperatures $T<30$~K (Fig.~\ref{fig:optical}(c)).

\textit{Discussion.---}
The non-equilibrium transport measurements probe transport under the presence of various electrical disturbances.  The ac voltage driving on the left side of the channel (Fig.~\ref{fig:nonlocal-Hall}(a)) produce a non-equilibrium condition that affects transport characteristics on the right side of the channel.    
%These quasiparticles need to be driven by the overall source-drain voltage $V_{sd}$, and the conductance of the channel on the right, which is spatially separated from the excitation region on the left, can detect the presence of quasiparticles through changes in conductance.  An
The increase in conductance in the region sampled by $V_1$, brought about by increasing values of $V_{ac}$, can be interpreted as resulting from quasiparticles excited by $V_{ac}$ that drift into the area between electrodes 3 and 4.  Furthermore, the asymmetry of $I(V_1)$ yields insight into the charge of these carriers.  For values of $V_{ac}$ exceeding ~100 mV, the conductance for $V_1<0$ sharply exceeds that of its mirror opposite $V_1>0$.  The implication is that the conductance is larger when negatively charged carriers drift from the left side of the channel to the right side of the channel.

The photocurrent experiments demonstrate the ability of light to produce mobile charge carriers in the pseudogap phase.  The optical experiments can be interpreted in a consistent fashion, where the source of the pair breaking arises from light rather than electrical gating.  The onset of this photoconductive state coincides in temperature with other transport signatures of the pseudogap phase, and it seems clear that photoexcitation is creating quasiparticles that are immobilized by the pseudogap, and that these carriers live long enough to contribute excess current under suitable voltage bias conditions.  The threshold under which photocurrent is observed ($V_{dc}\approx300$~mV) is comparable to the threshold voltage measured in dc transport (Fig.~\ref{fig:diagram}(g)), consistent with a scenario that mobile carriers are first photoexcited and subsequently inhibited from recombining by the applied dc bias.  The temperature-dependence of the photocurrent provides a signature that the origin of the mobile carriers is linked to the pseudogap phase.

Collectively, the electrical and optical experiments demonstrate that the insulating pseudogap phase is unstable to perturbations that can produce long-lived negatively charged carriers.  An obvious interpretation is that these carriers are indeed
pre-formed electron pairs, similar to those that have been reported for
LaAlO$_3$/SrTiO$_3$ heterostructures and nanostructures \cite{Richter2013-lr,Cheng2015-fv, Annadi2018-az}. It is therefore plausible that SrTiO$_3$-based quantum well pseudogap behavior arises from electron pairing in the SmTiO$_3$/SrTiO$_3$/SmTiO$_3$ quantum wells.  What is unusual is the energy scale, which is at least an order of magnitude larger for the narrow SrTiO$_3$ quantum well.

The experiments described here do not directly point to electron pairing as the origin of the pseudogap phase.  They are consistent with this scenario, and suggest follow-up experiments that can more definitively identify the origin of the pseudogap behavior.  For example, mesoscopic devices such as single-electron transistors may be able to discern the charge of carriers in the paired and unpaired state.  Being able to confirm electron pairing as the origin of the pseudogap in this system would be a remarkable finding, because of the strong connection between quantum confinement and pseudogap energy.  Electron pairing at high energies, combined with phase coherence, can potentially lead to the development of engineered structures with unprecedented superconducting transition temperatures.

\begin{acknowledgments}
J.L. acknowledges support from NSF PHY-1913034 and NSF DMR-2225888. S.S. thanks the U.S. Department of Energy for support (Award No. DE-SC0020305).
\end{acknowledgments}

\end{document}